\documentstyle[preprint,aps,prc,floats,epsfig]{revtex}
\tighten
\let\jnfont=\rm
\def\NPB#1,{{\jnfont Nucl.\ Phys.\ }{\bf B#1},}
\def\PLB#1,{{\jnfont Phys.\ Lett.\ B }{\bf #1},}
\def\PRD#1,{{\jnfont Phys.\ Rev.\ D }{\bf #1},}
\def\PRL#1,{{\jnfont Phys.\ Rev.\ Lett.\ }{\bf #1},}
\def\ZPC#1,{{\jnfont Z.~Phys.\ C }{\bf #1},}

\begin{document}
\draft
\preprint{}

\preprint{\vbox{\hbox{\bf hep-ph/0504???} }}
                                                                                        
\title{Some studies on dark energy related problems}
                                                                                        
\author{\ \\[2mm] Fei Wang$^1$, Jin Min Yang$^{2,1}$ }
                                                                                        
\address{ \ \\[2mm]
$^1$ {\it Institute of Theoretical Physics, Academia Sinica,
          Beijing 100080, China } \\
$^2$ {\it CCAST (World Laboratory), P. O. Box 8730, Beijing 100080, China }}
                                                                                        
\maketitle
                                                                                        
\begin{abstract}

In this work we perform some studies related to dark energy. Firstly, 
we propose a dynamical approach to explain the dark energy content of the universe. 
We assume that a massless scalar field couples to the Hubble parameter with some 
Planck-mass suppressed interactions. This scalar field developes 
a Hubble parameter-dependent (thus time-dependent) vacum expectation value, which 
renders a time-independent relative density for dark energy and thus can explain the 
coincidence of the dark energy density of the universe. 
Secondly, we assume the dark matter particle is meta-stable
and decays very lately into the dark energy scalar field. Such a conversion of matter to dark 
energy can give an explanation for the starting time of the accelerating expansion of the universe.
Thirdly, we introduce multiple Affleck-Dine fields to the landscape scenario of dark
energy in order to have the required baryon-asymmetry universe.  
  
\end{abstract}
                                                                                        

\section{Introduction}
The nature of the content of the universe is a great mystery in today's physical science. 
The Wilkinson Microwave Anisotropy Probe (WMAP) collaboration gives fairly accurate values
on the content of the universe \cite{wmap} 
\begin{eqnarray}
       \Omega_m  =  0.27_{-0.04}^{+0.04} \ , ~
       \Omega_b =  0.044_{-0.004}^{+0.004} \ ,~
       \Omega_{\Lambda}=0.73_{-0.04}^{+0.04} \ , ~
       \eta=(6.14\pm0.25)\times10^{-10} \ ,
\end{eqnarray}
where $\Omega_m$, $\Omega_b$ and $\Omega_{\Lambda}$ denotes the density of total matter, baryonic 
matter and dark energy, respectively.  $\eta$ denotes the baryon to photon ratio.   
We see that, coincidentally, the dark matter density is comparable to the dark energy density as 
well as to the baryonic matter density. Such a coincidence needs to be understood.

For the explanation of such a coincidence, some phenomenologically dynamical approaches have been 
proposed, such as the quintessence \cite{quint1,quint2}, phantom \cite{phantom} and k-essence 
\cite{essence}. In this note we propose a new dynamical approach to explain the dark energy 
coincidence in the content of the universe. In this approach a massless scalar field is assumed 
to couple to the Hubble parameter with some Planck-mass suppressed interactions. This scalar 
field developes a Hubble parameter-dependent (thus time-dependent) vacum expectation value, which 
renders time-independent dark energy density and thus can explain the coincidence of the dark 
energy density of the universe. 
We further assume the dark matter particle is meta-stable
and decays very lately into the dark energy scalar field. Such a conversion of matter to dark energy
can give an explanation for another puzzle, namely the starting time of the accelerating expansion 
of the universe.

Another puzzle related to the dark energy is the smallness of dark energy (or cosmological constant).
Weinberg used anthropic principle \cite{weinberg} to argue that fine-tunning is needed by the existence 
of human beings. Such an approach is based on the hypothesis of multiple vacua, each of which has identical 
physicical properties but different value of vacuum energy. 
Motivated by such an approach, landscape from string theory 
is used to provide the vast amount of vacuum. In such a landscape scenario, the vast amount ($\sim 10^{120}$) 
of vacua of the potential arise from a large number of fields (say 100 $\sim$ 300), which ensure the 
statistical selection to give a plausible vacuum energy. However, the baryon content may be over-washed 
out by sphaleron effects even at temperature moderatly beneath $\Lambda_{QCD}$ and give a small baryon 
to photon raito to be consistent with baryon-symmetry universe \cite{hamed}. In order to give the required 
baryon-asymmetry universe, we propose to use multiple Affleck-Dine fields. We find that with  
multiple Affleck-Dine fields the baryon content can be much higher, which has a large range to ensure 
the present asymmetric baryon abundance.

This article is organized as follows. In Sec. II we elucidate the new dynamical approach to dark energy.
In Sec. III we discuss the conversion of dark matter to dark energy, which cause the accelerating expansion 
of the universe. In Sec. III we introduce multiple Affleck-Dine fields to the  landscape scenario in order
to  give the required baryon-asymmetry universe. The conclusion is given in Sec. V.  

\section{Dynamical approach to dark energy} 

The smallness of the dark energy (cosmological constant) may imply the existence of some new 
fundamental law of nature. It is posssible for dark energy to have dynamical behaviors.
We know that dark energy can be related to the Hubble constant $H_0$ and Planck 
scale $M_{pl}$ by a see-saw mechnism \footnote{Our universe can be described by the 
Robertson-Walker metric
\begin{eqnarray*}
ds^2=dt^2-a^2(t)\left(\frac{dr^2}{1-kr^2}+r^2d{\theta}^2+r^2{\sin}^2{\theta}d{\varphi}^2 \right)
\end{eqnarray*}
Here $a(t)$ is the scale factor, $k$ can be chosen to be $+1$,$-1$, or $0$ for spaces of constant 
positive, negative or zero curvature, respectively.
The see-saw relation can also be seen in the Friedman equation
\begin{eqnarray*}
  H^2=\frac{8 \pi G}{3}  \left( \rho_m+\rho_{DE} \right) .
\end{eqnarray*}
The gravitional constant is related to the planck scale by $G= {\hbar} c^5/M_{pl}^2 $. 
$H$ is the Hubble parameter defined by $H(t)=\dot{a(t)}/a(t)$ where $t=t_0$ gives 
the Hubble constant. }
\begin{equation}
\frac{\Lambda}{H_0}\sim \frac{M_{pl}}{\Lambda} ,
\end{equation}
where $\Lambda$ is related to the dark energy density by $\rho_{DE}\sim{\Lambda}^4$.
We can attribute the varying of the dark energy to a massless scalar field $\phi$. 
Such a massless scalar can be the Nambu-Goldstone boson from the broken of the global $U(1)$ 
R-symmetry \cite{r-symmetry} by gravity effects. We can phenomenologically adopt the potential 
of the form \cite{dine} 
\begin{equation}
V(\phi) =H^2 {\phi}^2 f\left(\frac{{\phi}^2}{M_{pl}^2}\right) .
\end{equation}
It is quite possible for the second derivative of the potential to be negative.
As an effective theory, the flatness of the potential for the massless scalar can be lifted by 
higher order gravitional force. We introduce the Planck scale-suppressed terms which preserve 
$-\phi\leftrightarrow\phi$ symmetry 
\begin{equation}
V\left(\phi\right)\approx -H^2 {\phi}^2+\frac{\lambda}{M^2_{pl}}{\phi}^6 ,
\end{equation}
where $\lambda$ is a dimensionless constant or variable of ${\cal O}(1)$,
characterizing the coupling of $\phi$. 

The vacum expectation value of  $\phi$ is then given by 
\begin{equation}
\langle \phi \rangle^4 \sim \frac{1}{3 \lambda}H^2(t) M_{pl}^2 .
\end{equation}
Here we can see two features for our approach:
\begin{itemize}
\item[(1)]  The dark energy density $\rho_{DE} \sim \langle \phi \rangle^4$
            is  time-denpendent, which makes the relevant density  
            $\Omega_{\Lambda}=\rho_{DE}/(\rho_m+\rho_{DE})$ time-independent.             
\item[(2)] When $t=t_0$ (present time),  $\rho_{DE} \sim \langle \phi \rangle^4$ can naturally take 
           the required value $\sim H_0^2 M_{pl}^2$.
\end{itemize}
So in this way, the dark energy coincidence in the content of the universe can be understood. 

We also note that the coincidence of dark matter density and baryonic matter density
can be understood in the Affleck-Dine mechanism for baryogenesis.    
In this mechanism, baryonic number can be generated dynamically. The oscillation of 
the field is generically unstable with spatial perturbation and can condenses into 
non-topological solitons called Q-balls \cite{enqvist,kasuya,fujii,ichikawa}. 
The late decays of these Q-balls into dark matter relate baryonic matter density to  
dark matter density \cite{wang,jedamzik}.
 
\section{Conversion of dark matter to dark energy}

For the dynamical scalar field $\phi$ introduced in the preceding section, we can also
introduce some subdominate terms for its interaction with the dark matter particle,
which is assumed to be a scalar $\tilde f$ (say the super-partner of sterile neutrino)  
\begin{equation}
\frac{1}{M_{pl}^3}{\phi}^6 {\tilde f } .
\end{equation}
It will not cause any phenomenological problems in particle physics   
since it is much suppressed. Through this interaction the scalar $\tilde f$ decays into  $\phi$
and its lifetime $\tau$ can be estimated as 
\begin{equation}
\tau^{-1}\sim {\left(\frac{m_{\tilde f}}{M_{pl}}\right)}^6 m_{\tilde f}.
\end{equation}
Suppose $m_{\tilde f}$ is large as the $\sim 10^9$ GeV, such decay occurs 
at the time scale
\begin{equation}
\tau  \sim 10^{18} s ,
\end{equation}
which is of the order of the age of the universe. Thus such meta-stable particles can be a
component of the relic dark matter (for some extensive studies on the cosmology of
meta-stable sfermions, see \cite{feng}).  

Through such decays, dark matter particles are being converted to dark energy field particles,
which can explain the starting time ( $z\sim1$ ) of the accelerating expansion of the universe, 
as explained in the following.

From the Friedman equation
\begin{equation}
\frac{\ddot{a}(t)}{a(t)}=-\frac{4 \pi G}{3} \left( \rho+3 P \right)
\end{equation}
we know that the accelerating expansion of the universe starts when $ \rho+3 P $ becomes negative.
Here  $\rho=\rho_{m}+\rho_{DE}$ is the total energy density and $P$ is the presure.   
The equation of state is given by 
\begin{equation}
\omega=\frac{P}{\rho}=\frac{T-V}{T+V}
\end{equation} 
where T,V are the kinematic and potential energy, respectively.
$\omega$ is vanishing for matter. Generally, the equation of state for dark energy is similar to that of 
the quintessence model and $\omega > -1$.
The decay of the dark matter can alter the state of equation for dark energy by kinematic terms.

Then we get 
\begin{eqnarray}
\rho+3 P = \rho_{m}+\rho_{DE}+3 \left( P_{m}+P_{DE}\right)=\rho_{m}+(1+3\omega)\rho_{DE} .
\end{eqnarray}
As the decay of $\tilde f$ into $\phi$ proceeds, $\rho_{m}$ is getting smaller and $\rho_{DE}$ 
is getting larger, and at some point $\rho_{m}+(1+3\omega)\rho_{DE}$ becomes negative since 
$1+3\omega$ is negative.  Such a point is called the critical point, at which  
$\rho_{m}+\rho_{DE}=\Omega \rho_{critical}$ ($\Omega\equiv \rho/\rho_{critical}$) and
\footnote{At the critical point, if we naively use $\omega=-1$, we find that the dark energy 
constitutes about $1/3$ of the total conent of the universe. However,such a portion can be 
increased when  $\omega$ is larger than $-1$, which is highly justified.}
\begin{eqnarray}
\frac{\rho_{DE}}{\Omega \rho_{critical}}=\frac{1}{-3\omega} \, .
\end{eqnarray}
Since in our scenario such a critical point happens as a result of the decay of
the meta-stable dark matter particle and the decay occurs at the time scale of
$ 10^{18} s$, we get an understanding why the universe starts accelerating expansion 
quite lately  ( $z\sim1$).

Note that in our scenario the dynamical field $\phi$ may couple to graviton. The radiation of
gravitons can slowly decrease the kinematic energy. So the transition of dark matter to dark 
energy cause the slow loss of universe content. In this way our scenario predicts that the universe 
is evolving toward an Anti-de-Sitter universe.  If the universe is flat till now ($\Omega=1$), 
it will evolve to be open ($\Omega<1$).

\section{Multiple Affleck-Dine fields in landscape}
 
In Affleck-Dine mechanism, a complex scalar field has U(1) symmetry, which  correponds to a conserved 
current and is regarded as baryon number. It has potential interactions that violate CP. It can develop 
a large vacuum expectation value and when oscillation begins, it can give a net baryon number. 
Supersymmetry provides the natural candidates for such scalar fields. The vast number of 
flat directions\cite{flat} that carry baryon or lepton number can have vanishing quartic terms.
Nonrenormalizable higher-dimensional terms can lift the flat directions which then can give a large 
vacuum expectation value. Here we propose to use multiple flat directions (each of which denoted by $\Phi_i$) 
to generate the net baryon number in our universe.

We consider a superpotential, which can lift such flat directions from supersymmetry breaking terms,  
to have the following leading form
\begin{equation}
 W_n^i=\frac{1}{M^n} {\Phi_{i}}^{n+3} ,
\end{equation}
where $M$ is the scale of new physics and $n$ is some integer.    
 So the corresponding potential takes the form      
\begin{equation}
 V= -H^2 {\left|\Phi_{i}\right|}^2+\frac{1}{M^{2 n}} {\left|\Phi_{i}\right|}^{2 n+4} .
\end{equation} 
The leading sources of $B$ and $CP$ violations come from supersymmetry breaking terms (by gravity)
\begin{equation}  
 a m_{3/2} W_n^i+b H W_n^i ,
\end{equation}
where $a$ and $b$ are complex dimensionless constants and $m_{3/2}$ is the gravitino mass.
The relative phase in these two terms, $\delta= {\tan}^{-1} ( a b^*/ | a b|  ) $, violates $CP$.
We can chose $n$ and $a,b$ to ensure each potential to have several meta-stable vacuum expectation 
values with very different magnitudes as illustrated in Fig. 1 (Acceptable selection in natural consideration 
may require that the magnitudes be different by $10^3\sim 10^4$). 
\begin{figure}
\begin{center}
\includegraphics[width=10cm,height=8cm,angle=0]{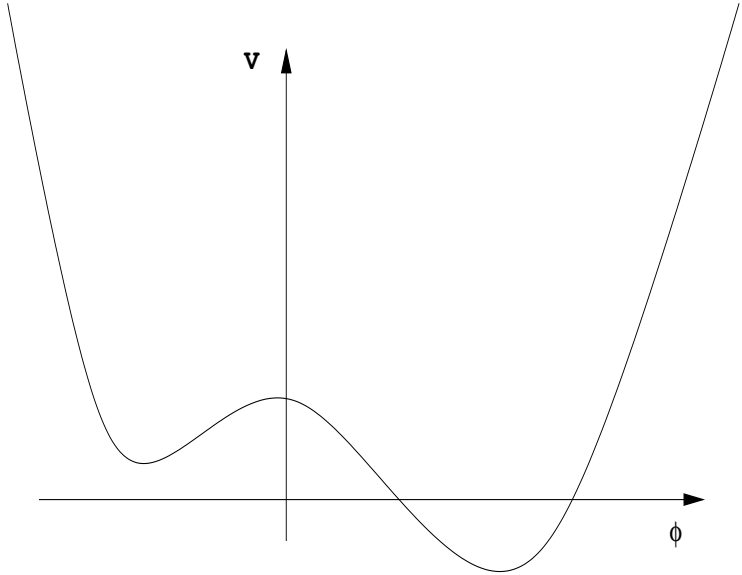}
\end{center}
\caption{\it The illustration plot of different vacuum expectation values in the flat direction potential 
         which can generate different baryonic contents.}
\end{figure}
The two vacuum expectation values of $\Phi_i$ are given as 
\begin{equation}
\Phi_{i,0}\approx M {\left( \frac{H}{M}\right)}^{1/(n+1)}
\end{equation}
and  
\begin{equation}
     \Phi_{i,0}\approx M {\left( \frac{2 [Re(a) m_{3/2}+Re(b) H] }{M}\right)}^{1/(n+1)} .
\end{equation}
So we can get more than $10^{120}$ vacua for $100 \sim 300$ Affleck-Dine fields.

The evolution of the baryon number is
\begin{equation}
\frac{d n_{B}^i}{d t}=\frac{ \sin(\delta) m_{3/2}}{M^n} {\Phi}_{i}^{n+3} \, .
\end{equation}
Naive estimation gives ( we assume $H \sim 1/t $):
\begin{equation}
n_{B}=\sum\limits_{i}{} \frac{ \sin(\delta)}{M^n} {\Phi}_{i,0}^{n+3} .
\end{equation}
As each of the two metastable vacua differs significantly, the combination of multiple fields can give 
a large range for baryonic content. The rate of washing out baryon asymmetry is given by \cite{hamed}
\begin{equation}
\frac{d n_{B}}{d t}=-\Gamma . 
\end{equation}
Here $\Gamma$ is given by
\begin{equation}
\Gamma={\alpha_{W}}^4 T {\left(\frac{M_{W}(T)}{{\alpha}_{W} T}\right)}^7 e^{-\frac{M_W(T)}{{\alpha}_W T}} ,
\end{equation}
where at zero temperature $M_{W}$ is given by
\begin{equation}
M_{W}\sim g_{W} f\sim g_{W} \frac{\Lambda_{QCD}}{4 \pi} .
\end{equation}
The residue abundance in our multiple fields case can be several orders higher
\footnote{$n_{total}\sim n_{field}\times n_{differ}$ can be as higher as $10^3\sim10^{3(n+3)}$. 
Here $n_{field}$ is the number of Affleck-Dine fields and $n_{differ}$ is the order of difference
between the meta-stable vacuum values.} than ordinary approach, which can greatly enhance the residue 
value for baryon content and thus make it possible to be consistent with baryon-asymmetry universe.

\section{Conclusion} 
 We performed some studies related to dark energy.
Firstly, we proposed a dynamical approach to explain the dark energy content of the universe. 
We assumed that a massless scalar field couples to the Hubble parameter with some 
Planck-mass suppressed interactions. Such a scalar field developes 
a Hubble parameter-dependent (thus time-dependent) vacum expectation value, which 
renders a time-independent relative density for dark energy and thus can explain the 
coincidence of the dark energy density of the universe. 
Secondly, we assumed the dark matter particle is meta-stable
and decays very lately into the dark energy scalar field. Such a conversion of matter to dark 
energy can give an explanation for the starting time of the accelerating expansion of the universe.
Finally, we introduced multiple Affleck-Dine fields to the landscape scenario of dark
energy in order to have the required baryon-asymmetry universe.    

\section*{Acknowledgement}
We are grateful to Dr Zongkuang Guo,Dr Wei Hao and Dr Dingfang Zeng for enlightment discussions. 
This work is supported in part by National Natural Science Foundation of China.

\end{document}